\documentstyle[prd,aps,amssymb,newlfont,epsfig]{revtex}

\begin{document}
\draft
\preprint{
\begin{tabular}{r}
   UWThPh-1999-32
\\ DFTT 25/99
\\ hep-ph/9905246
\\
\end{tabular}
}
\title{Sterile neutrinos?}
\author{S.M. Bilenky}
\address{Joint Institute for Nuclear Research, Dubna, Russia, and\\
Institute for Theoretical Physics, University of Vienna,\\
Boltzmanngasse 5, A--1090 Vienna, Austria}
\author{C. Giunti}
\address{INFN, Sez. di Torino, and Dip. di Fisica Teorica,
Univ. di Torino,\\
Via P. Giuria 1, I--10125 Torino, Italy}
\maketitle
\begin{abstract}
The notion of sterile neutrinos is discussed.
The schemes of mixing
of four massive neutrinos, which imply the existence of sterile neutrinos,
are briefly considered.
Several model independent methods that allow to 
reveal possible transitions of solar neutrinos into sterile states are
presented.
\end{abstract}
\pacs{Talk presented by S.M. Bilenky at the
VIII International Workshop on \textit{Neutrino Telescopes},
February 23--26, 1999, Venice.
UWThPh-1999-32, DFTT 25/99, hep-ph/9905246.}  

\section{Introduction}
\label{Introduction}

The notion of sterile neutrinos was introduced by B. Pontecorvo in 1967
\cite{Pontecorvo67}.
In the last years the possibility of transitions of flavor
neutrinos into noninteracting sterile states has been widely discussed in the
literature.
This interest in sterile neutrinos is connected mainly with
the result of the LSND experiment \cite{LSND}
that together with the results of atmospheric \cite{atm-exp,SK-atm} and
solar \cite{sun-exp,SK-sun} neutrino experiments imply the existence of sterile
neutrinos \cite{four,BGG-AB,Okada-Yasuda-97,BGGS-98-BBN,six}.

\section{Notion of sterile neutrinos}
\label{Notion of sterile neutrinos}

Let us start with the discussion of the possibilities of sterile
neutrinos to appear in neutrino mixing schemes
(see, for example,
Refs. \cite{Bilenky-Pontecorvo-78,Bilenky-Petcov-87,BGG-review}).
Flavor neutrinos $\nu_e$, $\nu_{\mu}$, $\nu_{\tau}$,
are determined by the standard charged-current (CC) and
neutral-current (NC) weak interactions
\begin{eqnarray}
\mathcal{L}_I^{\mathrm{CC}}
=
-\frac{g}{2\sqrt{2}} \ j_{\alpha}^{\mathrm{CC}} \, W^\alpha
+
\mathrm{h.c.}
\,,
& \qquad &
j_{\alpha}^{\mathrm{CC}}
=
2 \sum_{\ell=e,\mu,\tau} \overline{\nu_{\ell L}} \, \gamma_\alpha \, \ell_L
\,,
\label{CC}
\\
\mathcal{L}_I^{\mathrm{NC}}
=
-\frac{g}{2\cos{\theta_W}} \ j_{\alpha}^{\mathrm{NC}} \, Z^\alpha
\,,
& \qquad &
j_{\alpha}^{\mathrm{NC}}
=
\sum_{\ell=e,\mu,\tau} \overline{\nu_{\ell L}} \, \gamma_\alpha \, \nu_{\ell L}
\,,
\label{NC}
\end{eqnarray}
that conserve the electron $L_e$, muon $L_{\mu}$ and tau $L_{\tau}$ 
lepton numbers,
\begin{equation}
\sum{L_e} = \mathrm{const}
\,,
\qquad
\sum{L_\mu} = \mathrm{const}
\,,
\qquad
\sum{L_\tau} = \mathrm{const}
\,.
\label{lepton numbers}
\end{equation}

If there is neutrino mixing, the conservation laws (\ref{lepton numbers})
are violated.
A \textit{neutrino mass term} that does not conserve lepton numbers
has the general form
\begin{equation}
\mathcal{L}^M = - \overline{n_R} \, M \, n_L + \mathrm{h.c.}
\,,
\end{equation}
where $n_L$ and $n_R$ are $N$-component columns
and $M$ is a $N{\times}N$ matrix.
There are two different possibilities for $n_L$:

\begin{enumerate}

\item
Only flavor fields $\nu_\ell$ ($\ell=e,\mu,\tau$) enter into $n_L$:
\begin{equation}
n_L
=
\left( \begin{array}{c}
\nu_{eL}\\ 
\nu_{\mu L}\\
\nu_{\tau L}
\end{array} \right)
\,. 
\end{equation}
In this case $M$ is a $3\times3$ matrix and for the mixing we have
\begin{equation}
\nu_{\ell L} = \sum_i U_{\ell i} \, \nu_{iL}
\qquad
(\ell=e,\mu,\tau)
\,,
\label{3mixing}
\end{equation}
where $U$ is a unitary matrix and $\nu_i$ is the field of neutrinos with
mass $m_i$.
The nature of neutrinos $\nu_i$ with definite mass
depends on $n_{R}$:

\begin{enumerate}

\item
If
\begin{equation}
n_R
=
\left( \begin{array}{c}
\nu_{eR}\\
\nu_{\mu R}\\
\tau_{\tau R}
\end{array} \right)
\,,
\end{equation}
where $\nu_{\ell R}$
($\ell=e,\mu,\tau$)
are right-handed neutrino fields,
the total lepton number
$ L = L_{e} + L_{\mu} + L_{\tau} $
is conserved and the neutrinos $\nu_i$ with definite mass
are \textit{Dirac particles}.
The corresponding mass term is called
``Dirac mass term''.

\item
If
\begin{equation}
n_R
=
\left( \begin{array}{c}
(\nu_{eL})^c\\
(\nu_{\mu L})^c\\
(\nu_{\tau L})^c
\end{array} \right)
\,,
\end{equation}
where
$ (\nu_{\ell L})^c = \mathcal{C} \, \overline{\nu_{\ell L}}^T $
is the charge-conjugated (right-handed)
component of the left-handed field
$\nu_{\ell L}$
($\ell=e,\mu,\tau$ and
$\mathcal{C}$ is the matrix of charge conjugation),
there are no
conserved lepton numbers and neutrinos with definite masses 
are \textit{Majorana particles} ($\nu_i^c=\nu_i$).
The corresponding mass term is called ``Majorana mass term''.

\end{enumerate}

From Eq. (\ref{3mixing}) it follows that in both cases only transitions between
active neutrinos
$ \nu_\ell \leftrightarrows \nu_{\ell'} $
are possible.
Notice that
with the investigation of neutrino oscillations it is not possible to
distinguish the case of Dirac neutrinos from the case of Majorana
neutrinos \cite{Dirac-Majorana}.

\item
In the column $n_L$ not only the flavor fields $\nu_{\ell L}$
($\ell=e,\mu,\tau$)
but also
other fields $\nu_{sL}$ ($s=s_1,s_2,\ldots$)
enter 
\begin{equation}
n_L
=
\left( \begin{array}{c}
\nu_{eL}\\
\nu_{\mu L}\\
\nu_{\tau L}\\
\nu_{s_1 L}\\
\vdots
\end{array} \right)
\,.
\end{equation}
The fields
$\nu_{sL}$ do not enter in the standard CC and NC weak interactions
(\ref{CC}), (\ref{NC}) and are called ``sterile''.
It is possible that there are three sterile
fields $\nu_{sL}$ that are the charge-conjugated components of
right-handed neutrino fields, $\nu_{sL}=(\nu_{sR})^c$ ($s=e,\mu,\tau$).
However,
the sterile fields could also be fields of some other particles (SUSY, \ldots).

In this case,
for the neutrino mixing we have
\begin{equation}
\nu_{\alpha L}
=
\sum_{i=1}^{3+N_s}
U_{\alpha i} \, \nu_{iL}
\,.
\label{mixing}
\end{equation}
where $N_s$ is the number of sterile fields and 
$\nu_i$ ($i=1,\ldots,3+N_s$) is the field of neutrinos with mass $m_i$.
Let us stress that the number of sterile
fields depends on the concrete scheme of neutrino mixing.

The nature of neutrinos with definite mass depends on $n_R$.
If $n_R = (n_L)^c$, the $\nu_i$'s
are Majorana particles.
If
$\nu_{sL}=(\nu_{sR})^c$ ($s=e,\mu,\tau$),
the corresponding mass term is called
``Dirac-Majorana mass term''.

If all the masses $m_i$ of the fields $\nu_i$
in Eq.(\ref{mixing}) are small,
not only transitions between flavor neutrinos
$ \nu_\ell \leftrightarrows \nu_{\ell'} $
but also transitions between flavor and sterile neutrinos
$ \nu_\ell \leftrightarrows \nu_s $
are possible. 

\end{enumerate}

\section{Possible ways to reveal the existence of the sterile neutrinos}
\label{Possible ways to reveal the existence of the sterile neutrinos}

There are two possible ways to reveal the existence of sterile neutrinos:

\begin{enumerate}

\item
Through the determination of the number of massive neutrinos.
If this number
is larger than the number of flavor neutrinos (three)
sterile neutrinos must exist.

\item
Through the measurement of the total transition probability of neutrino 
with definite flavor
($\nu_e$ or $\nu_{\mu}$) into all possible flavor neutrinos,
$ \sum_{\ell'=e,\mu,\tau} P_{\nu_\ell\to\nu_{\ell'}} $
($\ell=e,\mu$).
This probability can be 
determined from the investigation of
NC induced neutrino processes.
If
\begin{equation}
\sum_{\ell'=e,\mu,\tau} P_{\nu_\ell\to\nu_{\ell'}}
<
1
\qquad
(\ell=e,\mu)
\,,
\label{disappearance}
\end{equation}
from the unitarity of the mixing matrix it follows that
active flavor neutrinos transform into sterile states.

\end{enumerate}

\section{Schemes of mixing of four massive neutrinos}
\label{Schemes of mixing of four massive neutrinos}

The data of neutrino oscillation experiments
indicate the existence of three different scales of
neutrino mass-squared difference
$\Delta{m}^2$:

\begin{enumerate}

\item
$ \Delta{m}^2_{\mathrm{atm}} \sim 10^{-3} \, \mathrm{eV}^2 $
from atmospheric neutrino experiments \cite{atm-exp,SK-atm};

\item
$ \Delta{m}^2_{\mathrm{sun}} \sim 10^{-5} \, \mathrm{eV}^2 $
(or $1 0^{-10} \, \mathrm{eV}^2 $)
from solar neutrino experiments
\cite{sun-exp,SK-sun,BKS-98-sun-analysis,SK-99-sun-analysis}; 

\item
$ \Delta{m}^2_{\mathrm{LSND}} \sim 1 \, \mathrm{eV}^2 $
from the LSND experiment \cite{LSND}.

\end{enumerate}

Three different scales of
$\Delta{m}^2$ can be accommodated only if
at least four massive neutrinos exist in nature
\cite{four,BGG-AB,Okada-Yasuda-97,BGGS-98-BBN,six}.
In other words, the existing data indicate that
sterile neutrinos exist. 

In the framework of the minimal scheme with four massive neutrinos,
from the existing data it follows \cite{Okada-Yasuda-97,BGGS-98-BBN}
that the dominant transitions of solar neutrinos
are $\nu_e \to \nu_s$.
Below we will present the corresponding arguments.

\begin{figure}[h]
\begin{center}
\setlength{\unitlength}{1.0cm}
\begin{tabular*}{0.99\linewidth}{@{\extracolsep{\fill}}cccccc}
\begin{picture}(1,4) 
\thicklines
\put(0.1,0.2){\vector(0,1){3.8}}
\put(0.0,0.2){\line(1,0){0.2}}
\put(0.4,0.15){\makebox(0,0)[l]{$m_1$}}
\put(0.0,0.4){\line(1,0){0.2}}
\put(0.4,0.45){\makebox(0,0)[l]{$m_2$}}
\put(0.0,0.8){\line(1,0){0.2}}
\put(0.4,0.8){\makebox(0,0)[l]{$m_3$}}
\put(0.0,3.5){\line(1,0){0.2}}
\put(0.4,3.5){\makebox(0,0)[l]{$m_4$}}
\end{picture}
&
\begin{picture}(1,4) 
\thicklines
\put(0.1,0.2){\vector(0,1){3.8}}
\put(0.0,0.2){\line(1,0){0.2}}
\put(0.4,0.2){\makebox(0,0)[l]{$m_1$}}
\put(0.0,0.6){\line(1,0){0.2}}
\put(0.4,0.55){\makebox(0,0)[l]{$m_2$}}
\put(0.0,0.8){\line(1,0){0.2}}
\put(0.4,0.85){\makebox(0,0)[l]{$m_3$}}
\put(0.0,3.5){\line(1,0){0.2}}
\put(0.4,3.5){\makebox(0,0)[l]{$m_4$}}
\end{picture}
&
\begin{picture}(1,4) 
\thicklines
\put(0.1,0.2){\vector(0,1){3.8}}
\put(0.0,0.2){\line(1,0){0.2}}
\put(0.4,0.2){\makebox(0,0)[l]{$m_1$}}
\put(0.0,2.9){\line(1,0){0.2}}
\put(0.4,2.9){\makebox(0,0)[l]{$m_2$}}
\put(0.0,3.3){\line(1,0){0.2}}
\put(0.4,3.25){\makebox(0,0)[l]{$m_3$}}
\put(0.0,3.5){\line(1,0){0.2}}
\put(0.4,3.55){\makebox(0,0)[l]{$m_4$}}
\end{picture}
&
\begin{picture}(1,4) 
\thicklines
\put(0.1,0.2){\vector(0,1){3.8}}
\put(0.0,0.2){\line(1,0){0.2}}
\put(0.4,0.2){\makebox(0,0)[l]{$m_1$}}
\put(0.0,2.9){\line(1,0){0.2}}
\put(0.4,2.85){\makebox(0,0)[l]{$m_2$}}
\put(0.0,3.1){\line(1,0){0.2}}
\put(0.4,3.15){\makebox(0,0)[l]{$m_3$}}
\put(0.0,3.5){\line(1,0){0.2}}
\put(0.4,3.5){\makebox(0,0)[l]{$m_4$}}
\end{picture}
&
\begin{picture}(1,4) 
\thicklines
\put(0.1,0.2){\vector(0,1){3.8}}
\put(0.0,0.2){\line(1,0){0.2}}
\put(0.4,0.2){\makebox(0,0)[l]{$m_1$}}
\put(0.0,0.6){\line(1,0){0.2}}
\put(0.4,0.6){\makebox(0,0)[l]{$m_2$}}
\put(0.0,3.3){\line(1,0){0.2}}
\put(0.4,3.25){\makebox(0,0)[l]{$m_3$}}
\put(0.0,3.5){\line(1,0){0.2}}
\put(0.4,3.55){\makebox(0,0)[l]{$m_4$}}
\end{picture}
&
\begin{picture}(1,4) 
\thicklines
\put(0.1,0.2){\vector(0,1){3.8}}
\put(0.0,0.2){\line(1,0){0.2}}
\put(0.4,0.15){\makebox(0,0)[l]{$m_1$}}
\put(0.0,0.4){\line(1,0){0.2}}
\put(0.4,0.45){\makebox(0,0)[l]{$m_2$}}
\put(0.0,3.1){\line(1,0){0.2}}
\put(0.4,3.1){\makebox(0,0)[l]{$m_3$}}
\put(0.0,3.5){\line(1,0){0.2}}
\put(0.4,3.5){\makebox(0,0)[l]{$m_4$}}
\end{picture}
\\
(I) & (II) & (III) & (IV) & (A) & (B)
\end{tabular*}
\end{center}
\caption{ \label{4spectra} \small
The six types of neutrino mass spectra that can accommodate 
the solar, atmospheric and LSND scales of $\Delta{m}^2$.
The different
distances between the masses on the vertical axes symbolize the
different scales of $\Delta{m}^2$.
Class 1 is constituted by the spectra (I)--(IV),
whereas class 2 comprises the spectra (A) and (B).}
\end{figure}
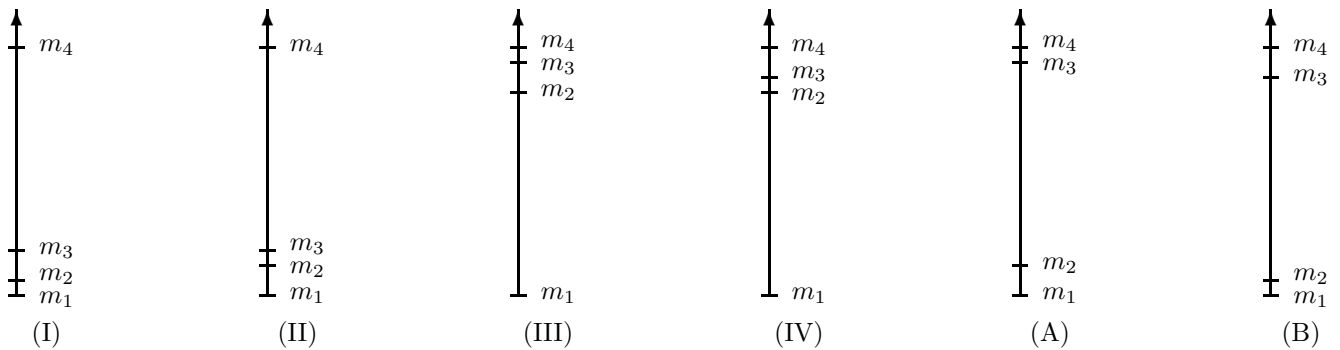

Figure \ref{4spectra}
shows the six types of spectra of four massive neutrinos
that can accommodate the
solar, atmospheric and LSND ranges of neutrino mass-squared differences.
These spectra are divided in two classes:
class 1 constituted by the spectra (I)--(IV)
and class 2 comprising the spectra (A) and (B).
In the spectra of class 1 a group of three close masses is
separated from the fourth mass by the LSND gap of $ \sim 1 \, \mathrm{eV}^2 $.
In the spectra
of class 2 two pairs of close masses are separated by the LSND gap.

In the case of the spectra of class 1
the amplitude of
$\nu_{\mu} \to \nu_e$ transitions
in short-baseline (SBL)
experiments
is strongly suppressed and the upper
bound of the corresponding oscillation amplitude
obtained from the exclusion curves of SBL $\bar\nu_e$ and $\nu_\mu$
disappearance experiments, is smaller than the value of the amplitude
of $\nu_\mu \to \nu_e$ oscillations found in the LSND
experiment \cite{BGG-AB}.
Thus, the schemes of mixing of
four massive neutrinos belonging to class 1 are not
compatible with the existing neutrino oscillation data.

Only the schemes of mixing of four massive neutrinos with mass
spectra of class 2,
\begin{equation}
\mbox{(A)}
\qquad
\underbrace{
\overbrace{m_1 < m_2}^{\mathrm{atm}}
\ll
\overbrace{m_3 < m_4}^{\mathrm{sun}}
}_{\mathrm{LSND}}
\qquad \mbox{and} \qquad
\mbox{(B)}
\qquad
\underbrace{
\overbrace{m_1 < m_2}^{\mathrm{sun}}
\ll
\overbrace{m_3 < m_4}^{\mathrm{atm}}
}_{\mathrm{LSND}}
\,,
\label{AB}
\end{equation}
can describe all the existing
neutrino oscillation data. 
In this case,
the largest mass squared difference
$\Delta{m}^2_{41} \equiv m_4^2 - m_1^2$ is relevant for the
oscillations in SBL reactor and accelerator experiments.
The corresponding transition probabilities are \cite{BGKP}
\begin{eqnarray}
&&
P_{\nu_\alpha\to\nu_\beta}
=
\frac{1}{2} \, A_{\alpha;\beta} \,
\left( 1 - \cos \frac{ \Delta{m}^2_{41} L }{ 2 p } \right)
\,,
\label{Ptran}
\\
&&
P_{\nu_\alpha\to\nu_\alpha}
=
1
-
\frac{1}{2} \, B_{\alpha;\alpha} \,
\left( 1 - \cos \frac{ \Delta{m}^2_{41} L }{ 2 p } \right)
\,.
\label{Psurv}
\end{eqnarray}
Here
$L$ is the distance between neutrino source and neutrino detector,
$p$ is neutrino momentum,
and
$A_{\alpha;\beta}$,
$B_{\alpha;\alpha}$
are the oscillation amplitudes given by
\begin{eqnarray}
&&
A_{\alpha;\beta}
=
4
\left|
\sum_i
U_{{\beta}i}
\,
U_{{\alpha}i}^*
\right|^2
\,,
\label{Aab}
\\
&&
B_{\alpha;\alpha}
=
4
\left( \sum_i |U_{{\alpha}i}|^2 \right)
\left( 1 - \sum_i |U_{{\alpha}i}|^2 \right)
\,,
\label{Baa}
\end{eqnarray}
where the index $i$ runs over the values $1,2$
or $3,4$.

The quantities
$ \sum_i |U_{\alpha i}|^2 $
for
$\alpha=e,\mu$
are constrained by the results of SBL reactor $\bar\nu_e\to\bar\nu_e$
and accelerator $\nu_{\mu}\to\nu_{\mu}$ disappearance experiments
in which no indications in favor of neutrino oscillations were found:
\begin{equation}
\sum_i |U_{\alpha i}|^2
\leq
a^0_\alpha
\qquad \mbox{or} \qquad
1 - \sum_i |U_{\alpha i}|^2
\leq
a^0_\alpha
\,.
\label{a0alpha}
\end{equation}
Here 
\begin{equation}
a_\alpha^0
=
\frac{1}{2}
\left( 1 - \sqrt{ 1 - B_{\alpha;\alpha}^0 } \, \right)
\,.
\label{aa0}
\end{equation}
where
$B_{\alpha;\alpha}^0$
is the upper bound of the amplitude of 
$\nu_\alpha\to\nu_\alpha$ transitions
obtained from the exclusion curves of
SBL reactor and accelerator disappearance experiments.
Using the exclusion plots obtained in the
reactor Bugey \cite{Bugey-95} experiment
and in the accelerator CDHS \cite{CDHS-84} and CCFR \cite{CCFR-84}
experiments we have
\cite{BBGK}
\begin{eqnarray}
&&
a^0_e \lesssim 4 \times 10^{-2}
\qquad \mbox{for} \qquad
\Delta{m}^2_{41} \gtrsim 4 \times 10^{-2} \, \mathrm{eV}^2
\,,
\label{a0_e}
\\
&&
a_{\mu}^0 \lesssim 2 \times 10^{-1}
\qquad \mbox{for} \qquad
\Delta{m}^2_{41} \gtrsim 3 \times 10^{-1} \, \mathrm{eV}^2
\,.
\label{a0_mu}
\end{eqnarray}

Further constraints on the quantities
$ \sum_i |U_{\alpha i}|^2 $
for
$\alpha=e,\mu$
can be found by taking
into account the results of solar and atmospheric neutrino experiments.
The survival
probabilities of solar $\nu_e$'s and atmospheric $\nu_{\mu}$'s
are constrained by
(see \cite{BGKP,BGG-AB,BGG-review})
\begin{equation}
P_{\nu_e\to\nu_e}^{\mathrm{sun}}
\geq
\sum_i |U_{\alpha i}|^4
\,,
\qquad
P_{\nu_\mu\to\nu_\mu}^{\mathrm{atm}}
\geq
\left( 1 - \sum_i |U_{\alpha i}|^2 \right)^2
\,,
\label{Pmin}
\end{equation}
where the index $i$ runs over 1,2 in scheme A and over 3,4 in scheme B.

Let us introduce the quantities
\begin{equation}
c_\alpha
\equiv
\sum_i |U_{\alpha i}|^2
\qquad
(\alpha=e,\mu,\tau,s)
\,,
\label{c_alpha}
\end{equation}
with the index $i$
running over 1,2 in scheme A and over 3,4 in scheme B.

From
Eqs. (\ref{a0_e}), (\ref{a0_mu})
and
(\ref{Pmin})
we conclude \cite{BGG-AB}
that the four-neutrino schemes A and B
are compatible with the results of solar and atmospheric neutrino experiments
only for
\begin{equation}
c_e \leq a^0_e
\qquad \mbox{and} \qquad
1 - c_\mu \leq a^0_\mu
\,.
\label{R2}
\end{equation}

Constraints on the elements of the mixing matrix $U$ can be also obtained
from the limit on the effective number of neutrinos $N_\nu$
in Big-Bang Nucleosynthesis (BBN)
(see, for example, \cite{Schramm-Turner-98}). 
The analysis of recent data yields the upper bound $N_\nu \leq 3.2$
at 95\% CL \cite{Burles-99},
which implies that \cite{Okada-Yasuda-97,BGGS-98-BBN}
\begin{equation}
c_s \ll 1
\,.
\label{c_s}
\end{equation}
Taking now into account the unitarity relation
\begin{equation}
\sum_{\alpha=e,\mu,\tau,s} c_\alpha = 2
\,,
\label{unitarity}
\end{equation}
from Eqs. (\ref{R2}) and (\ref{c_s}) we come to the conclusion that
$c_e$ and $c_s$ are small and $c_{\mu}$ and $c_{\tau}$ are large.
In the approximation
\begin{equation}
c_e \ll 1
\,,
\qquad
c_s \ll 1
\,,
\qquad
c_\mu \simeq 1
\,,
\qquad
c_\tau \simeq 1
\,,
\label{approx}
\end{equation}
in scheme A we have
\begin{equation}
\begin{array}{l} \displaystyle
\nu_{eL}
=
\cos \vartheta \, \nu_{3L}
+
\sin \vartheta \, \nu_{4L}
\,,
\\ \displaystyle
\nu_{sL}
=
- \sin \vartheta \, \nu_{3L}
+
\cos \vartheta \, \nu_{4L}
\,,
\\ \displaystyle
\nu_{\mu L}
=
\cos \gamma \, \nu_{1L}
+
\sin \gamma \, \nu_{2L}
\,,
\\ \displaystyle
\nu_{\tau L}
=
- \sin \gamma \, \nu_{1L}
+
\cos \gamma \, \nu_{2L}
\,.
\end{array}
\label{approx-mix}
\end{equation}
where $\vartheta$ and $\gamma$ are mixing angles.
The corresponding mixing 
relations in scheme B can be obtained from Eq. (\ref{approx-mix}) with the change
$1,2\leftrightarrows3,4$.
From Eq. (\ref{approx-mix}) one can see that the dominant transitions of solar
neutrinos are $\nu_e\to\nu_s$ and
the dominant transitions of
atmospheric neutrinos and neutrinos in long-baseline (LBL) experiments are
$\nu_\mu\to\nu_\tau$. 

\section{Possible tests for $\lowercase{\nu_e\to\nu_s}$ transitions of solar neutrinos}
\label{Possible tests}

Here we will consider possible model independent methods of searching for 
$\nu_e\to\nu_s$ transitions
in future solar neutrino experiments \cite{BG-sterile-sun}.

In the SNO experiment \cite{SNO}
solar neutrinos will be detected through the observation of the CC reaction
\begin{equation}
\nu_e + d \to e^- + p + p
\,,
\label{SNO-CC}
\end{equation}
of the NC reaction
\begin{equation}
\nu + d \to \nu + p + n
\,,
\label{SNO-NC}
\end{equation}
and of the elastic-scattering (ES) reaction
\begin{equation}
\nu + e^- \to \nu + e^-
\,.
\label{SNO-ES}
\end{equation}
Because of the large energy thresholds
($ \sim 5 \, \mathrm{MeV} $ for the CC and ES processes and 2.2 MeV
for the NC process),
mainly $^8$B neutrinos will be detected in the SNO experiment.
Let us write the initial flux of $^8$B neutrinos 
as a function of energy $E$ in the form
\begin{equation}
\phi_{^8\mathrm{B}}(E)
=
X(E) \, \Phi_{^8\mathrm{B}}
\,.
\label{phi}
\end{equation}
Here
$\Phi_{^8\mathrm{B}}$
is the total flux and
$X(E)$
is a known function
$ \left( \int X(E) \, \mathrm{d}E = 1 \right)$
that characterizes the spectrum of $\nu_e$'s in the decay
$ {}^8\mathrm{B} \to {}^8\mathrm{Be} + e^+ + \nu_e $.
The NC event rate is given by
\begin{equation}
N_{\mathrm{NC}}
=
\langle
\sum_{\ell=e,\mu,\tau}
P_{\nu_e\to\nu_\ell}
\rangle_{{\nu}d}
\
\langle \sigma_{{\nu}d}^{\mathrm{NC}} \rangle
\
\Phi_{^8\mathrm{B}}
\,.
\label{NC event rate}
\end{equation}
Here
$
\langle
\sum_{\ell=e,\mu,\tau}
P_{\nu_e\to\nu_\ell}
\rangle_{{\nu}d}
$
is the average value of the total transition probability
of solar $\nu_e$'s into all possible flavor neutrinos
and
\begin{equation}
\langle \sigma_{{\nu}d}^{\mathrm{NC}} \rangle
\simeq
4.7 \times 10^{-43} \, \mathrm{cm}^2
\label{sigma-NC}
\end{equation}
is the average value of the cross section of the NC process (\ref{SNO-NC}). 

It is obvious that
$
\langle
\sum_{\ell=e,\mu,\tau}
P_{\nu_e\to\nu_\ell}
\rangle_{{\nu}d}
<
1
$
if $\nu_e\to\nu_s$ transitions take place.
We cannot, however, determine
the quantity
$
\langle
\sum_{\ell=e,\mu,\tau}
P_{\nu_e\to\nu_\ell}
\rangle_{{\nu}d}
$
from (\ref{NC event rate})
in a model independent way,
because knowledge of the total flux
$\Phi_{^8\mathrm{B}}$
is needed.

Information on the total flux
$\Phi_{^8\mathrm{B}}$
can be obtained from the data of the Super-Kamiokande experiment.
Indeed, we have
\begin{equation}
\Sigma_{{\nu}e}
=
\langle
\sum_{\ell=e,\mu,\tau}
P_{\nu_e\to\nu_\ell}
\rangle_{{\nu}e}
\
\langle \sigma_{{\nu_\mu}e} \rangle
\
\Phi_{^8\mathrm{B}}
\,,
\label{Sigma1}
\end{equation}
where 
\begin{equation}
\langle \sigma_{{\nu_\mu}e} \rangle
\simeq
2 \times 10^{-45} \, \mathrm{cm}^2
\label{sigma-nu_mu-e}
\end{equation}
is the average cross section of $\nu_{\mu}$-$e$ scattering and
\begin{equation}
\Sigma_{{\nu}e}
=
N_{{\nu}e}
-
\int_{E_{\mathrm{th}}}
\left(
\sigma_{{\nu_e}e}
-
\sigma_{{\nu_\mu}e}
\right)
\phi_{\nu_e}(E)
\
\mathrm{d}E
\,,
\label{Sigma2}
\end{equation}
is the NC contribution to the
$\nu_e e \to \nu_e e$ event rate $N_{{\nu}e}$.
In Eq. (\ref{Sigma2})
$\sigma_{{\nu_e}e}$
($\sigma_{{\nu_\mu}e}$)
is the cross section of $\nu_e$-$e$ ($\nu_\mu$-$e$) scattering and
$\phi_{\nu_e}(E)$
is the flux of solar $\nu_e$'s with energy $E$ on the earth,
that will be measured in the SNO experiment
through the observation of the CC reaction (\ref{SNO-CC}).
Combining the relations (\ref{NC event rate}) and (\ref{Sigma1})
we obtain the ratio
\begin{equation}
R
=
\frac
{ \Sigma_{{\nu}e} \ \langle \sigma_{{\nu}d}^{\mathrm{NC}} \rangle }
{ N_{\mathrm{NC}} \ \langle \sigma_{{\nu_\mu}e} \rangle }
=
\frac
{
\langle
\sum_{\ell=e,\mu,\tau}
P_{\nu_e\to\nu_\ell}
\rangle_{{\nu}e}
}
{
\langle
\sum_{\ell=e,\mu,\tau}
P_{\nu_e\to\nu_\ell}
\rangle_{{\nu}d}
}
\,,
\label{R}
\end{equation}
that is independent from the
the total flux
$\Phi_{^8\mathrm{B}}$.
Only measurable (and known) quantities enter in the ratio $R$.
From Eq. (\ref{R}) it is obvious that $R\neq1$
only if
$
\sum_{\ell=e,\mu,\tau}
P_{\nu_e\to\nu_\ell}
<
1
$,
\textit{i.e.} if solar neutrinos transfer
into sterile states.
Let us notice, however, that the ratio $R$ can be different from
one only if the probability
$
\sum_{\ell=e,\mu,\tau}
P_{\nu_e\to\nu_\ell}
$
depends on the neutrino energy.

\begin{figure}[h]
\begin{center}
\mbox{\epsfig{file=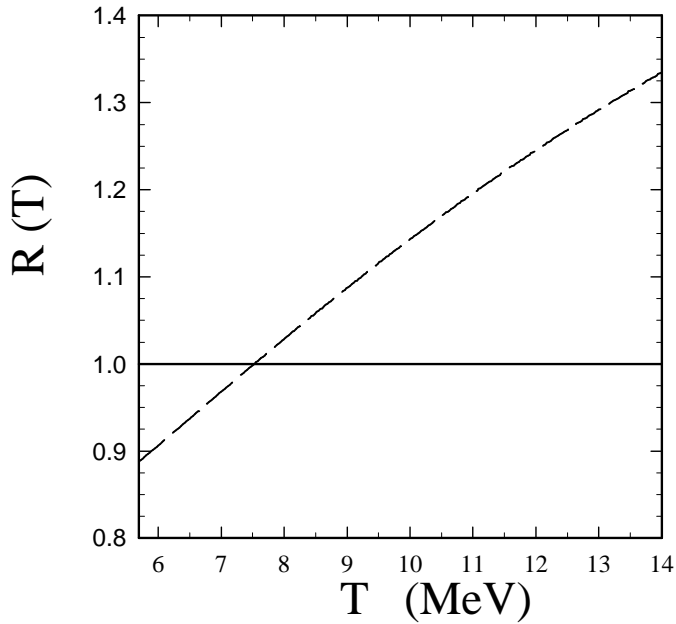,width=0.5\linewidth}}
\end{center}
\caption{ \label{reses} \small
Ratio
$R(T)$
(see Eq.(\protect\ref{R(T)}))
calculated under the assumption of
$\nu_{e}$--$\nu_{\mathrm{s}}$
mixing with the values of the mixing parameters
$ \Delta{m}^2 = 4.5 \times 10^{-6} \, \mathrm{eV}^2 $,
$ \sin^2 2 \vartheta = 7.0 \times 10^{-3} $
(the small mixing angle MSW solution).}
\end{figure}

Another possibility to reveal
transitions of solar neutrinos into sterile states could be realized
by combining the
Super-Kamiokande recoil electron spectrum and the spectrum of
$\nu_e$'s on the earth that will be determined in the SNO experiment.
We have
\begin{equation}
\Sigma_{{\nu}e}(T)
=
\langle
\sum_{\ell=e,\mu,\tau}
P_{\nu_e\to\nu_\ell}
\rangle_{{\nu}e,T}
\
\langle \frac{ \mathrm{d}\sigma_{{\nu_\mu}e} }{ \mathrm{d}T } \rangle
\
\Phi_{^8\mathrm{B}}
\,,
\label{Sigma3}
\end{equation}
where $T$ is the recoil electron kinetic energy in the Super-Kamiokande experiment,
\begin{equation}
\Sigma_{{\nu}e}(T)
=
N_{{\nu}e}(T)
-
\int_{E_{\mathrm{min}}(T)}
\left(
\frac{ \mathrm{d}\sigma_{{\nu_e}e} }{ \mathrm{d}T }
-
\frac{ \mathrm{d}\sigma_{{\nu_\mu}e} }{ \mathrm{d}T }
\right)
\phi_{\nu_e}(E)
\
\mathrm{d}E
\,,
\label{Sigma4}
\end{equation}
and
$
E_{\mathrm{min}}(T)
=
\frac{T}{2}
\,
\sqrt{ 1 + \frac{2m_e}{T} }
$.
From relation (\ref{Sigma3}) it is obvious that if the quantity 
\begin{equation}
R_{\mathrm{ES}}(T)
=
\Sigma_{{\nu}e}(T)
\Big/
\langle \frac{ \mathrm{d}\sigma_{{\nu_\mu}e} }{ \mathrm{d}T } \rangle
\label{RES(T)}
\end{equation}
depends on $T$ it means that the probability
$
P_{\nu_e\to\nu_s}
=
1
-
\sum_{\ell=e,\mu,\tau}
P_{\nu_e\to\nu_\ell}
$
is different from zero
and depends on neutrino energy.
The dependence on $T$ of the ratio
\begin{equation}
R(T)
=
\frac{
\Sigma_{{\nu}e}(T)
\
\langle \sigma_{{\nu_\mu}e} \rangle
}
{
\Sigma_{{\nu}e}
\
\langle \frac{ \mathrm{d}\sigma_{{\nu_\mu}e} }{ \mathrm{d}T } \rangle
}
\label{R(T)}
\end{equation}
is presented in Fig. \ref{reses}.
The dashed line corresponds to the small mixing angle MSW solution of the solar neutrino
problem
in the case of
$\nu_e\to\nu_s$ transitions
with
$ \Delta{m}^2 = 4.5 \times 10^{-6} \, \mathrm{eV}^2 $,
$ \sin^2 2 \vartheta = 7.0 \times 10^{-3} $.

The most direct way to search for 
$\nu_e\to\nu_s$ transitions
of atmospheric neutrinos is to investigate the NC process
\begin{equation}
\nu + N \to \nu + \pi^0 + X
\,.
\label{atm-NC}
\end{equation}
This possibility has been discussed in Ref. \cite{atm-NC}.

\section{Conclusions}
\label{Conclusions}

If we accept all the existing indications in favor of neutrino
oscillations we come to the conclusion that sterile neutrinos
must exist.
Thus, from the phenomenological point of view the problem of
sterile neutrinos is connected with the correctness of the existing 
indications in favor of neutrino oscillations
and first of all with the correctness of the indications obtained
in the LSND experiment, 
the only accelerator SBL experiment in which
neutrino oscillations have been observed.  

Through the investigation of NC processes it is possible to search for
transitions of active neutrinos into sterile states in solar, atmospheric
SBL and LBL experiments.
Several possible tests are based only on the unitarity of the
mixing matrix and can be done in a model independent way.

\acknowledgements

S.M.B. would like to thank the Institute for Theoretical Physics of
the University of Vienna for its hospitality.

\end{document}